\documentclass[pra,aps,showpacs,twocolumn,epsfig]{revtex4}
 \usepackage{graphicx}
 \begin{document}
 \title{ Mesoscopic Superposition of States with Sub-Planck Structures in Phase Space }
 \author{G. S. Agarwal and P. K. Pathak}
 \address{Physical Research Laboratory,
 Navrangpura, Ahmedabad-380 009, India}
 \date{\today}
 \begin{abstract}
  We propose cavity quantum electrodynamics method, using
  dispersive interaction between atoms and a high quality
  cavity to realize the mesoscopic
  superposition of coherent states which would exhibit sub-Planck
  structures in phase space, {\it i.e.} the structures at a scale smaller than the Plank's
  constant ($\hbar$). These structures are direct signatures of quantum coherence
  and are formed as a result of interference
  between the two superposed cat states. In particular we focus on a
  superposition involving four coherent states.
  We show interferences in the conditional
  measurements involving two atoms.
 \end{abstract}
 \pacs{42.50.Gy, 32.80.Qk}
 \maketitle
\section{Introduction}
 In recent times mesoscopic superposition of states has attracted
 a great deal of attention as these superpositions exhibit very important
 interference effects \cite{SCHLEICH,KNIGHT,HAROCHE} many of which have now been realized
 experimentally \cite{AUFFEVES,BRUNE,RAIMOND,WINELAND,STROUD}. The simplest superposition
 would consist of two coherent states one centered at $\alpha$ and the other at
 $-\alpha$. Such a state is known to be an eigenstate of the operator $a^2$. It has
 been known that the passage of a single mode of the field in a coherent state through
 a Kerr medium could produce such a state \cite{TARA,TOMBESI,ENK}. In
 an earlier work \cite{TARA}, it was shown that a variety of
 other superpositions can be produced by a Kerr medium. In
 particular, one can produce eigenstates of the operator $
 a^4$. Such eigenstates are  superpositions of four coherent
 states. However an efficient production of such states would
 require large Kerr nonlinearity which is not available though some
 proposals for the enhancement of the Kerr nonlinearity exist
 \cite{IMAMOGLU}. The existence of such superpositions is closely
 connected to the occurrence of fractional revivals in the
 nonlinear dynamics of quantum systems \cite{STROUD,BANERJI,BANERJI2,STROUD2}.
 In particular, a fractional revival of order $1/4$ can produce a superposition of
 four coherent states. However instead of pursuing the production using Kerr medium,
 we propose to use cavity QED methods. We note that
 Haroche and  coworkers \cite{BRUNE,RAIMOND,HAROCHE} showed how cavity quantum
 electrodynamics  can be used to produce a superposition of two mesoscopic
 states. It turns out that one can have fairly large dispersive
 interaction in high quality cavities. This high dispersion has been
 utilized by several authors \cite{RP,GERRY} to
 produce a variety of entangled states and  nonclassical
  superpositions including a superposition of four coherent states.
 In this paper we show how to
 prepare superpositions of four coherent states by using
 resonant as well as dispersive interaction in a high quality cavity. This study is
 motivated by a recent finding of Zurek \cite{ZUREK} that a
 proper superposition of four coherent states which he refers
 to as a compass state, can exhibit regions in phase space with
 sub-Planck structures, {\it i. e.} the area of the variations of the
 two quadratures can be much smaller than $\hbar$. We demonstrate how the results of
 conditional measurements on three atoms passing in succession, through a high
 Q-cavity, can yield information on such a compass state.
 \section{COMPASS STATE FOR THE RADIATION FIELD}
 Consider a single mode radiation field specified by the annihilation and creation
operators $a$ and $a^{\dag}$. Let $|\alpha\rangle$ be a coherent
state for the field with amplitude $\alpha$. The most commonly
studied superpositions are of the form
 \begin{equation}
|\psi\rangle \sim |\alpha\rangle + |e^{i\theta}\alpha\rangle.
\label{cat}
 \end{equation}
 Here  $\theta$ is an arbitrary phase. Extensive literature on this state exists.
  It is well known \cite{SCHLEICH,KNIGHT}
 that the quantum character of this state is reflected in the regions of phase
 space  where the Wigner function becomes negative. The area of the negative region
is of the order of Planck constant. There are several methods of
producing such a state \cite{HAROCHE,WINELAND,STROUD,TARA}.
 Zurek \cite{ZUREK} has
studied a superposition state of  four Gaussian wave packets
\begin{equation}
\psi(x)\sim
exp\left\{-\frac{(x-x_{0})^2}{\xi^2}+\frac{ip_{0}x}{\hbar}\right\},
\end{equation}
with one each placed in the east, west, north and south direction
in the phase space and calculated the Wigner function for such a
state, defined by
\begin{equation}
W(x,p)=\frac{1}{2\pi\hbar}\int e^{ipy/\hbar}
\psi\left(x-\frac{y}{2}\right)\psi^{*}\left(x+\frac{y}{2}\right)dy.
\end{equation}
He found that it exhibits negative regions in phase space as well
as structures with areas which could be much smaller than Planck's
constant. Since coherent states correspond to Gaussian wave
packets, in the following we consider a superposition of four
coherent states of the form
\begin{equation}
|\phi\rangle = N\left(|\alpha\rangle + |i\alpha\rangle
+|-\alpha\rangle + |-i\alpha\rangle\right), \label{state}
 \end{equation}
 where $N$ is the normalization constant and $\alpha$ is complex.
The Wigner function for any state $|\phi\rangle$ can be obtained
using coherent states as \cite{AGARWAL}
\begin{eqnarray}
W(\gamma,\gamma^{*})=\frac{2}{\pi^{2}}e^{2|\gamma|^{2}}\int
\langle -\beta|\phi\rangle\langle\phi|\beta\rangle
e^{-2(\beta\gamma^{*}-\beta^{*}\gamma)}d^{2}\beta.
\label{defn}
\end{eqnarray}
 For
the state $(\ref{state})$ the Wigner function is found to be
\begin{widetext}
\begin{eqnarray}
W(\gamma,\gamma^{*})&=&|N|^2\frac{4e^{-2|\gamma|^2}}{\pi}
\left[2e^{-2|\alpha|^2}\cosh\left\{(1+i)\alpha\gamma^{*}+(1-i)\alpha^{*}\gamma\right\}
\cosh\left\{(1-i)\alpha\gamma^{*}+(1+i)\alpha^{*}\gamma\right\}\right.\nonumber\\
&&\left.+2\cos\left\{(1+i)\alpha\gamma^{*}+(1-i)\alpha^{*}\gamma\right\}
\cos\left\{(1-i)\alpha\gamma^{*}+(1+i)\alpha^{*}\gamma\right\}\right.\nonumber\\
&&\left.+e^{-\left(|\alpha|^2-(1+i)\alpha\gamma^{*}-(1-i)\alpha^{*}\gamma\right)}
\cos\left\{|\alpha|^2-(1+i)\alpha\gamma^{*}-(1-i)\alpha^{*}\gamma\right\}\right.\nonumber\\
&&\left.+e^{-\left(|\alpha|^2-(1-i)\alpha\gamma^{*}-(1+i)\alpha^{*}\gamma\right)}
\cos\left\{|\alpha|^2-(1-i)\alpha\gamma^{*}-(1+i)\alpha^{*}\gamma\right\}\right.\nonumber\\
&&\left.+e^{-\left(|\alpha|^2+(1+i)\alpha\gamma^{*}+(1-i)\alpha^{*}\gamma\right)}
\cos\left\{|\alpha|^2+(1+i)\alpha\gamma^{*}+(1-i)\alpha^{*}\gamma\right\}\right.\nonumber\\
&&\left.+e^{-\left(|\alpha|^2+(1-i)\alpha\gamma^{*}+(1+i)\alpha^{*}\gamma\right)}
\cos\left\{|\alpha|^2+(1-i)\alpha\gamma^{*}+(1+i)\alpha^{*}\gamma\right\}
\right]. \label{wigner}
\end{eqnarray}
\end{widetext}
Each cosine term in $(\ref{wigner})$ arises from the interference
of a pair of coherent states in the superposition state
$(\ref{state})$. The sub-Plank structures arise from further
interference of two cosine terms which come from the diagonal
pairs. The first two terms in $(\ref{wigner})$ are such terms
coming from the diagonal pairs $|\alpha\rangle$, $|-\alpha\rangle$
and $|i\alpha\rangle$, $|-i\alpha\rangle$. The first term is
significant for smaller values of $|\alpha|$ and  shows
exponential decrease  in the Wigner function away from the center
and the second term which is significant for larger values of
$|\alpha|$, shows the interference pattern in the central region
$(\gamma\rightarrow 0)$.
\begin{figure}
\centering
\caption{(Color online) The Wigner function for
mesoscopic superposition state
$N\left(|\alpha\rangle+|-\alpha\rangle+|i\alpha\rangle+|-i\alpha\rangle\right)$
for $|\alpha|=1$.}\label{fig1}
\end{figure}
\begin{figure}
\centering
\caption{(Color online) The Wigner function for
mesoscopic superposition state
$N\left(|\alpha\rangle+|-\alpha\rangle+|i\alpha\rangle+|-i\alpha\rangle\right)$
for $|\alpha|=5$.} \label{fig2}
\end{figure}
 In Figs \ref{fig1} and \ref{fig2}, we plot the Wigner function
 for some typical values of
 $|\alpha|$. We found that for smaller values of $|\alpha|$ (Fig\ref{fig1}),
 the central part has a continuum and no other
 structures appear but for larger values of $|\alpha|$ (Fig\ref{fig2}), a chess board
 pattern as noticed earlier by Zurek, appears in the central region.
 The reason for the disappearance of the
 interference pattern in the central region for smaller values of $|\alpha|$ is because
  in this case the coherent states overlap to a large extent so the
 interference effects are not visible.

 A natural question is how to produce the state
$(\ref{state})$. In what follows we show how the methods of cavity
quantum electrodynamics \cite{HAROCHE,BRUNE}can be generalized to
produce the state (\ref{state}).
\section{Generation of The Compass State Using dispersive interaction between atoms and cavity}
 Consider a single mode high Q-cavity containing a small
amount of a coherent field so that the initial state of the cavity
field is $|\alpha\rangle$. Let $\omega_c$ be the cavity frequency.
Consider the passage of a two level atom with the excited and
ground states $|e\rangle$ and $|g\rangle$ with transition
frequency $\omega$. The atom is initially prepared in a
superposition state
\begin{equation}
|\Phi\rangle = c_e|e\rangle + c_g|g\rangle.
\end{equation}
In a frame rotating with the atomic transition frequency $\omega$,
the interaction Hamiltonian is given by
\begin{equation}
H=\hbar\delta a^{\dag}a+ \hbar g(|e\rangle\langle g|a
+|g\rangle\langle e|a^{\dagger}),~~ \delta=(\omega_c-\omega).
\end{equation}
We assume that we are working in the dispersive limit so that
$\delta$ is large.  We can then do a second order perturbation
theory and obtain an effective Hamiltonian
\begin{equation}
H\simeq\hbar\delta a^{\dag}a+\phi_0\hbar
a^{\dagger}a|g\rangle\langle g| -\phi_0\hbar
aa^{\dagger}|e\rangle\langle e|, \label{ham}
\end{equation}
where the parameter $\phi_0$ is equal to ${g^{2}}/{\delta}$.
Physically it gives the shift of the excited state in the absence
of any cavity field. Under the effect of the Hamiltonian
$(\ref{ham})$, the states evolve as
\begin{eqnarray}
&&|g,n\rangle\rightarrow e^{-in\phi_0 \tau-in\delta\tau}|g,n\rangle\nonumber\\
&&|e,n\rangle\rightarrow e^{i(n+1)\phi_0
\tau-in\delta\tau}|e,n\rangle, \label{evol}
\end{eqnarray}
where $\tau$ is the interaction time. Using $(\ref{evol})$, we
easily obtain the evolution of a field in a coherent state
$|\alpha\rangle$
\begin{eqnarray}
&&|g,\alpha\rangle\rightarrow|g,\alpha e^{-i\phi
-i\delta\tau}\rangle\nonumber\\
&&|e,\alpha\rangle\rightarrow e^{i\phi}|e,\alpha e^{i\phi-i\delta
\tau}\rangle,~\phi=\phi_0\tau.
\end{eqnarray}
Therefore the atom field system in the state $|\Phi,\alpha\rangle$
will evolve into
\begin{eqnarray}
|\Phi,\alpha\rangle\rightarrow c_ee^{i\phi}|e,\alpha
e^{i\phi-i\delta \tau}\rangle+c_g|g,\alpha
e^{-i\phi-i\delta\tau}\rangle.
 \label{sup}
\end{eqnarray}
The probability of detection of the atom in the state $
|\psi\rangle=\psi_e|e\rangle+\psi_g|g\rangle$ will be
\begin{eqnarray}
P_{\psi}&=&||c_e\psi^*_ee^{i\phi}|\alpha e^{i\phi-i\delta
\tau}\rangle+c_g\psi_g^*|\alpha e^{-i\phi-i\delta\tau}\rangle||^2
\label{brac}\\
&=&|c_e\psi^*_e|^2+|c_g\psi^*_g|^2+\nonumber\\&&2 real
\left(c_g^*\psi_gc_e\psi^*_ee^{i\phi}\langle\alpha
e^{-i(\phi+\delta\tau)}|\alpha
e^{i(\phi-\delta\tau)}\rangle\right). \label{exp}
\end{eqnarray}
The last term in $(\ref{exp})$ yields the interference fringes.
For the special case of the initial state and the detection state
having equal superposition of the ground and the excited states
$|c_g^*\psi_gc_e\psi_e^*|=1/4$. The visibility depends on the
scalar product of two coherent states that are shifted in phase by
$2\phi$. The phase shift is a measure of the cavity interaction.
Haroche and coworkers have used the above for the production and
detection of mesoscopic superposition of the field states. In the
present case the generated mesoscopic superposition is the state
in Eq $(\ref{brac})$ under the $||~~||$ sign.

 We next demonstrate how the compass
state can be produced by following similar ideas. Let us write the
state $(\ref{sup})$ in the form
\begin{equation}
|\Phi,\alpha\rangle=f_e|e\rangle|\alpha_e\rangle+f_g|g\rangle|\alpha_g\rangle.
\label{10}
\end{equation}
Let us consider the passage of two atoms labeled as A and B in
succession through the cavity. After the passage of the atom A we
get the state $(\ref{10})$. Clearly the net state of the system
consisting of two atoms $A$, $B$ and the cavity field would have
the structure
\begin{eqnarray}
|\Psi\rangle=f_eh_e|e_A,e_B\rangle|\alpha_{ee^{'}}\rangle
+f_eh_g|e_A,g_B\rangle|\alpha_{eg^{'}}\rangle\nonumber\\
+f_gh_e|g_A,e_B\rangle|\alpha_{ge^{'}}\rangle
+f_gh_g|g_A,g_B\rangle|\alpha_{gg^{'}}\rangle . \label{11}
\end{eqnarray}
The joint detection of the atoms in the state
$|\chi\rangle\equiv\chi_{ee^{'}}|e_A,e_B\rangle+\chi_{eg^{'}}|e_A,g_B\rangle
+\chi_{ge^{'}}|g_A,e_B\rangle+\chi_{gg^{'}}|g_A,g_B\rangle$ will
project state$(\ref{11})$ to (unnormalized state)
\begin{eqnarray}
\langle\chi|\Psi\rangle\equiv|C\rangle=f_eh_e\chi_{ee^{'}}^*|\alpha_{ee^{'}}\rangle+
f_eh_g\chi_{eg^{'}}^*|\alpha_{eg^{'}}\rangle\nonumber\\
+f_gh_e\chi_{ge^{'}}^*|\alpha_{ge^{'}}\rangle+
f_gh_g\chi_{gg^{'}}^*|\alpha_{gg^{'}}\rangle . \label{12}
\end{eqnarray}
Clearly such a conditional detection reduces the state of the
cavity field to a state which in general would be a mesoscopic
superposition of four coherent sates $|\alpha_{ij}\rangle$. The
value of $\alpha_{ij}$ can be read from Eq $(\ref{sup})$:
\begin{eqnarray}
&&\alpha_{ee^{'}}=\alpha_0
e^{i\phi+i\phi^{'}},\alpha_{eg^{'}}=\alpha_0
e^{i\phi-i\phi^{'}}\nonumber\\
&&\alpha_{ge^{'}}=\alpha_0
e^{-i\phi+i\phi^{'}},\alpha_{gg^{'}}=\alpha_0
e^{-i\phi-i\phi^{'}};\\
&&\phi=\frac{g_A^2\tau_A}{\delta},
\phi^{'}=\frac{g_B^2\tau_B}{\delta};\nonumber\\
&&\alpha_0=\alpha e^{-i\delta \tau-i\delta\tau^{'}}.\nonumber
\label{13}
\end{eqnarray}
Clearly by varying $\phi$ and $\phi^{'}$ we can produce a variety
of superpositions. Consider for example $\phi=\pi/4$ and
$\phi^{'}=\pi/2$, then
\begin{eqnarray}
&&\alpha_{ee^{'}}=\alpha_0 e^{3i\pi/4},\alpha_{eg^{'}}=\alpha_0
e^{-i\pi/4}\nonumber\\
&&\alpha_{ge^{'}}=\alpha_0 e^{i\pi/4},\alpha_{gg^{'}}=\alpha_0
e^{-3i\pi/4}, \label{14}
\end{eqnarray}
so the state $(\ref{12})$ is a compass state. The expansion
coefficients in $(\ref{12})$ depend on the initial preparation of
the atoms $A$ and $B$ and the detection of these atoms.This is
usually done by using two Ramsey zones before and after the
cavity. Let us for simplicity assume that
\begin{eqnarray}
|\Phi_j\rangle&=&\frac{1}{\sqrt{2}}\left(e^{i\eta_j}|e\rangle+e^{i\theta_j}|g\rangle\right);
~j=A,B\nonumber\\
|\chi\rangle &=&|\Phi_A^{'}\rangle|\Phi_B^{'}\rangle, \label{15}
\end{eqnarray}
where $|\Phi_j^{'}\rangle$ is obtained from $|\Phi_j\rangle$ by
using $\eta_j\rightarrow\eta_j^{'}$ and
$\theta_j\rightarrow\theta_j^{'}$. Substituting values of
$\alpha_{ij}$ from $(\ref{14})$ we rewrite $(\ref{12})$ as
\begin{eqnarray}
\label{16}
&&|C\rangle=\frac{1}{4}\left(e^{i(\eta_1+\eta_2+3\pi/4)}|-\alpha\rangle
+e^{i(\eta_1+\theta_2+\pi/4)}|\alpha\rangle\right.\nonumber\\
&&\left.~~~~~~+e^{i(\theta_1+\eta_2+\pi/2)}|i\alpha\rangle
+e^{i(\theta_1+\theta_2)}|-i\alpha\rangle\right),\\
&&\eta_1=\eta_A-\eta_A^{'},\eta_2=\eta_B-\eta_B^{'},\theta_1=\theta_A-\theta_A^{'},
 \theta_2=\theta_B-\theta_B^{'},\nonumber
\end{eqnarray}
 we have also set
$\alpha_0=\alpha e^{i\pi/4}$. For $\theta_1=\eta_1+\pi/4$ and
$\theta_2=\eta_2+\pi/2$ the state (\ref{16}) becomes the compass
state (\ref{state})
\begin{eqnarray}
|C\rangle=\frac{1}{4}e^{i(\eta_1+\eta_2+3\pi/4)}\left(|-\alpha\rangle
+|\alpha\rangle +|i\alpha\rangle +|-i\alpha\rangle\right).
\label{phase}
\end{eqnarray}
 It is clear that the probability
of joint measurements on the atoms $A$ and $B$ would be
\begin{equation}
P=Tr_c\langle \chi|\psi\rangle\langle\psi|\chi\rangle
\end{equation}
where $Tr_c$ stands for tracing over the cavity field. Using
Eq(\ref{12}) and Eq(\ref{15}), we find the result
\begin{widetext}
\begin{eqnarray}
P=\frac{1}{4}+\frac{1}{8}
real\left(e^{i(\theta_2-\eta_2-\pi/2)}\langle-\alpha|\alpha\rangle+
e^{i(\theta_1-\eta_1-\pi/4)}\langle-\alpha|i\alpha\rangle+
e^{i(\theta_1+\theta_2-\eta_1-\eta_2-3\pi/4)}\langle-\alpha|-i\alpha\rangle
\right.\nonumber\\
\left.+e^{i(\theta_1-\eta_1-\pi/4)}\langle\alpha|-i\alpha\rangle
+e^{i(\theta_1+\eta_2-\eta_1-\theta_2+\pi/4)}\langle\alpha|i\alpha\rangle
+e^{i(\theta_2-\eta_2-\pi/2)}\langle
i\alpha|-i\alpha\rangle\right)\equiv \langle C|C\rangle.
\label{17}
\end{eqnarray}
\end{widetext}
In Fig\ref{fig3} we show $P$ as a function of phases of initial
atomic state and the detected atomic state for $|\alpha|=1$. These
interferences become less prominent for larger values of
$|\alpha|$.  The exact nature of interferences depends on the
choice of the phase factors $\eta_j$ and $\theta_j$.

In order to explore the characteristics of the state
(\ref{phase}), we have to bring a third atom $C$ and examine the
probability of its detection in a given state. This would be
similar to what was done in the experiment of Brune {\it et. al.}
\cite{BRUNE} to study the
 CAT state. Another possibility would involve a probe atom interacting
 resonantly with the prepared field in the cavity as the compass state
 (\ref{state}) involves photon number states  which are multiples of four.
 We discuss in the appendix $A$, the differences in the excitation
 probabilities for different states in the cavity.
\begin{figure}
\centering
\caption{(Color online) The probability $P$ for
$|\alpha|^2=1$ is plotted with phases of the initial atomic state
and the detected state. The scale along $x$ axis and $y$ axis is
in units of $\pi$.} \label{fig3}
\end{figure}

Following the work of Davidovich {\it et. al} \cite{HAROCHE} we
can examine the effect of detection efficiency on the preparation
of compass state. If one atom passes through the cavity undetected
it will leave the cavity either in its excited state $|e\rangle$
or in its ground state $|g\rangle$ in both the cases it will
produce phase shift in all the superposed coherent states equally
and as a result it can not affect the compass state except
rotating it in phase space. So in the experimental realization of
such mesoscopic superposition high efficiency detection is not
necessary. We relegate the details to the appendix $B$. We further
note that more complex homodyne methods like the ones used in Ref.
\cite{AUFFEVES} can be employed to probe the phase space
distributions associated with the compass state.

Next we consider effects of decoherence on the compass state
(\ref{state}). This can be done using the master equation
\begin{equation}
\dot{\rho}=-\frac{\kappa}{2}(a^{\dag}a\rho-2a\rho a^{\dag}+\rho
a^{\dag}a),
\end{equation}
where $\kappa$ is cavity field decay parameter and we assume that
the cavity is at zero temperature. For initial state (\ref{state})
we find the density matrix after time $t$
\begin{widetext}
\begin{eqnarray}
\rho(t)&=&|N|^2\left[|\alpha_t\rangle\langle\alpha_t|+|-\alpha_t\rangle\langle-\alpha_t|
+|i\alpha_t\rangle\langle
i\alpha_t|+|-i\alpha_t\rangle\langle-i\alpha_t|+
e^{-2|\alpha|^2(1-e^{-\kappa
t})}\left(|\alpha_t\rangle\langle-\alpha_t|+|-\alpha_t\rangle\langle\alpha_t|+
|i\alpha_t\rangle\langle-i\alpha_t|\right.\right.\nonumber\\
&&\left.\left.+|-i\alpha_t\rangle\langle i\alpha_t|\right)
+e^{-|\alpha|^2(1-i)(1-e^{-\kappa
t})}\left(|\alpha_t\rangle\langle i\alpha_t|+
|-i\alpha_t\rangle\langle\alpha_t|+|-\alpha_t\rangle\langle-i\alpha_t|
+|i\alpha_t\rangle\langle-\alpha_t|\right)\right.\nonumber\\
&&\left.+e^{-|\alpha|^2(1+i)(1-e^{-\kappa
t})}\left(|i\alpha_t\rangle\langle\alpha_t|+|\alpha_t\rangle\langle-i\alpha_t|
+|-i\alpha_t\rangle\langle-\alpha_t|+|-\alpha_t\rangle\langle
i\alpha_t|\right)\right];~~\alpha_t=\alpha e^{-\kappa t/2}.
\end{eqnarray}
\end{widetext}
 The coherence of
the superposition decays as $e^{-2|\alpha|^2(1-e^{-\kappa t})}$
which is  $e^{-2|\alpha|^2\kappa t}$ in the limit $\kappa t<<1$.
Thus the life time of the compass state will be $t_c/2|\alpha|^2$,
$t_c$ is life time of the cavity field. So the life time for
compass state is same as for a Schrodinger cat state (\ref{cat}).

\section{Conclusions}
We discussed the properties of the compass state for radiation
field as well as the methods of generating a compass state using
the dispersive atom cavity interaction. We showed that the central
interference pattern in the Wigner function for mesoscopic
superposition of cat states appears for larger values of
$|\alpha|$ and disappears for smaller values. The conditional
measurements enable one to study some aspects of the mesoscopic
superposition of coherent states. We have also discussed the
effects of decoherence on compass state as well as the effects of
nonunity detection efficiency in the preparation of the compass
state.

\appendix
\section{}
The compass state can be detected using the methods sensitive to
its field statistics. For the compass state photon distribution is
very special, it has number states having photon number in the
integral multiple of four. The state (\ref{state}) can be
expressed in terms of number states as follow
\begin{equation}
|\phi\rangle=N\sum_p\frac{\alpha^{4p}}{\sqrt{(4p)!}}e^{-|\alpha|^2/2}|4p\rangle,
\end{equation}
where $p$ ia an integer. We propose a simple method for detecting
 the compass state using a two level atom interacting resonantly
with the cavity field as a probe. The Hamiltonian in the
interaction picture is
\begin{equation}
H=\hbar g(|e\rangle\langle g|a+a^{\dag}|g\rangle\langle g|),
\label{probe} \end{equation}
 where all symbols have their earlier defined meanings. Using
 above interaction Hamiltonian we can calculate the probabilities of
 detection for the atom in its different states after passing through
 the cavity. The probabilities of detection if atom enters the
 cavity in its lower state $|g\rangle$ and detected in its state
 $|g\rangle$ and $|e\rangle$, $P_g^g$ and $P_g^e$ respectively are
 \begin{eqnarray}
 P_g^g=\sum_p|N\frac{\alpha^{4p}}{\sqrt{(4p)!}}e^{-|\alpha|^2/2}\cos(2gt\sqrt{p})|^2,\\
 P_g^e=\sum_p|N\frac{\alpha^{4p}}{\sqrt{(4p)!}}e^{-|\alpha|^2/2}\sin(2gt\sqrt{p})|^2.
 \end{eqnarray}
 \begin{figure}
\centering
\includegraphics[width=3.5in]{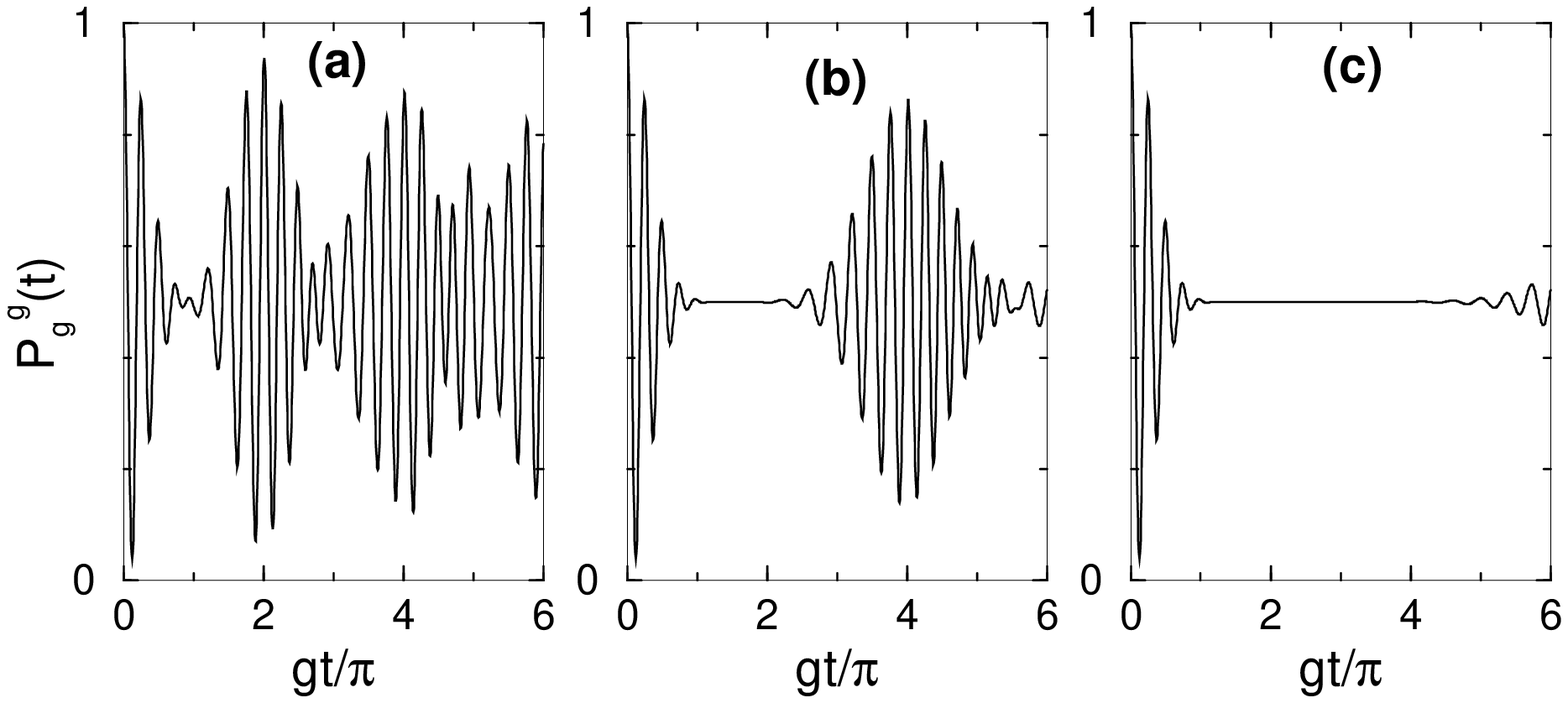}
\caption{The probability of detection of the atom in its ground
state $|g\rangle$ for cavity field in (a) compass state, (b)
 Schrodinger cat state $N_0(|\alpha\rangle+|-\alpha\rangle)$, and (c)
 coherent state $|\alpha\rangle$, for $|\alpha|^2=16$ .} \label{FIG4}
\end{figure}
 In the Fig\ref{FIG4} we show the comparison of detection
 probabilities for cavity field in compass state,
 Schrodinger cat state $N_0(|\alpha\rangle+|-\alpha\rangle)$, and
 coherent state $|\alpha\rangle$. We observe that the revival
 time is larger for cat state than the revival time for compass state
  and revival time is larger for coherent state than the revival
  time for cat states. The reduction in revival times is of
  increasing granular nature of photon distribution from coherent
  state to compass state.

\section{}
 In this appendix we follow the argument of Davidovich
{\it et. al.} \cite{HAROCHE} to show that the detection efficiency
is not a serious issue. After passing the first atom $A$ through
the cavity, the field state is projected
\begin{eqnarray}
|C_A\rangle=N\left(e^{i\eta_1+\pi/4}|\alpha
e^{i\pi/4}\rangle+e^{i\theta_1}|\alpha e^{-i\pi/4}\right),
\end{eqnarray}
where the velocity of atom $A$ is selected such that the phase
change in the cavity field $\phi=\pi/4$. If one atom similar to
atom $A$ passes through the cavity undetected, the combined state
will be
\begin{eqnarray}
|\psi'\rangle&=&Ne^{i(\eta_1+\pi/4)}\left(e^{i\eta_1+\pi/4}|\alpha'
e^{i\pi/2}\rangle+e^{i\theta_1}|\alpha'
\right)|e\rangle\nonumber\\
&+&Ne^{i\theta_1}\left(e^{i\eta_1+\pi/4}|\alpha'
\rangle+e^{i\theta_1}|\alpha'e^{-i\pi/2} \right)|g\rangle,
\end{eqnarray}
where $\alpha'=\alpha e^{-i\delta t_1}$. We trace out the atomic
state as the atom passes undetected, the cavity field will be in
the state
\begin{eqnarray}
|C'_A\rangle=N'\left[e^{i\eta_1+\pi/4}(|\alpha'\rangle+|i\alpha'\rangle)+e^{i\theta_1}
(|\alpha'\rangle+|-i\alpha'\rangle)\right].
\end{eqnarray}
Now if the second atom $B$ enters the cavity and detected after
passing the cavity in earlier defined states. The velocity of
second atom is chosen such that it changes phase of cavity field
by $\pi/2$. The detection of second atom will project the cavity
field in the state
\begin{eqnarray}
|C_B\rangle&=&\frac{N'}{2}\left[e^{i\eta_1+\eta_2+3\pi/4}(|\alpha''
e^{i\pi/2}\rangle+|i\alpha''
e^{i\pi/2})\right.\nonumber\\
&&\left.+e^{i(\theta_1+\eta_2+\pi/2)}(|\alpha''
e^{i\pi/2}\rangle+|-i\alpha''
e^{i\pi/2}\rangle)\right.\nonumber\\
&&\left.+e^{i\eta_1+\theta_2+\pi/4}(|\alpha''
e^{-i\pi/2}\rangle+|i\alpha''
e^{-i\pi/2})\right.\nonumber\\
&&\left.+e^{i(\theta_1+\theta_2)}(|\alpha''
e^{-i\pi/2}\rangle+|-i\alpha'' e^{-i\pi/2}\rangle)\right],
\label{det}
\end{eqnarray}
where $\alpha''=\alpha' e^{-i\delta t_2}$. For earlier defined
conditions on phases in the method for preparing the compass state
(\ref{state}), $\theta_1=\eta_1+\pi/4$ and $\theta_2=\eta_2+\pi/2$
the state (\ref{det}) becomes same as state (\ref{phase}). In a
similar way one can see that the prepared state will be a compass
state if one atom similar to atom $B$ passes undetected between
the atoms $A$ and $B$.
\end{document}